\documentclass[aps,prl,twocolumn,showpacs]{revtex4}
\usepackage{graphicx}

\def\8{\infty}
\def\oh{\frac{1}{2}}

\def\d{\partial}

\def\undertext#1{\vtop{\hbox{#1}\kern 1pt \hrule}}

\def\VEV#1{\langle\,#1\,\rangle}

\def\be{\begin{equation}}
\def\ee{\end{equation}}
\def\bea{\begin{eqnarray} & &}
\def\eea{\end{eqnarray}}

\def\rf#1{(\ref{#1})}

\def\rf#1{(\ref{#1})}

\def\rfs#1{Eq.~\rf{#1}}

\begin{document}


\title{Many body generalization of the Landau Zener problem}



\author{Alexander Altland}
\affiliation{Institut f\"ur Theoretische Physik,
Universit\"at zu K\"oln, Z\"ulpicher Str 77, 50937 K\"oln, Germany}

\author{V. Gurarie}

\affiliation{Department of Physics, University of Colorado,
Boulder CO 80309 USA}

\date{\today}

\begin{abstract}
  We formulate and approximately solve a specific many-body
  generalization of the Landau-Zener problem. Unlike with the single
  particle Landau-Zener problem, our system does not abide in the
  adiabatic ground state, even at very slow driving rates. The
  structure of the theory suggests that this finding reflects a more
  general phenomenon in the physics of adiabatically driven many
  particle systems. Our solution can be used to understand, for
  example, the behavior of two-level systems coupled to an
  electromagnetic field, as realized in cavity QED experiments.
 \end{abstract}

\maketitle

The Landau-Zener (LZ) problem describes a paradigmatic situation in
physics where two quantum levels cross each other in time.  In its
most basic form, the problem is represented by the Hamiltonian
\begin{equation} \label{eq:LZ}
H = \left( \matrix { \lambda t & g \cr g & - \lambda t} \right),
\end{equation}
where $t$ is time, $g$ the coupling constant, and $\lambda$  the
rate of change of the energy levels (here, as in the rest of the
paper, we set $\hbar =1$).  This Hamiltonian has two instantaneous
energy levels $E_{\pm} = \pm \sqrt{(\lambda t)^2 + g^2}$. Suppose in
the distant past, $t \rightarrow -\infty$, the system is in level
$E_-$. The goal then is to calculate the probability, $P$, to stay in
$E_-$ at $t \rightarrow +\infty$. Solving the corresponding time
dependent Schr\"odinger equation, Landau~ \cite{Landau1932} and
Zener~\cite{Zener1932} found 
\begin{equation} \label{eq:mLZ} 
P=1-e^{-\frac{\pi
      g^2}{\lambda}},
\end{equation}
as an exact answer to this question: if only the sweeping rate is slow
enough, it is exponentially likely that the system will abide in its
adiabatic ground state.  For 75 years, the Hamiltonian (\ref{eq:LZ}),
and its solution (\ref{eq:mLZ}) have been  used to
describe a huge spectrum of physical phenomena~\cite{LL}. Subsequent
generalizations of (\ref{eq:LZ}) include an extension to a
multi-channel environment wherein the 2 by 2 matrix is replaced by a
larger time-dependent matrix.  However, common to all those
problems~\cite{Demkov1966,Osherov1966,Kayanuma1985,Brundobler1992} is
that only a finite number of degrees of freedom participate in the
transition process (which manifests itself in transition probabilities
of the same algebraic structure as in Eq.~(\ref{eq:mLZ}).)

At the same time, there appears to be some interest in
genuine~\cite{fn_many_body} many-body generalizations of the LZ setup:
fundamentally, one would like to know whether a slowly driven many
body system will remain in its adiabatic ground state, in a manner
resembling the single particle case (\ref{eq:mLZ}). But there is also
applied relevance to the generalization. A number of existing
experimental setups provide a perspective to actually probe the
transition rates of a many-body LZ problem. Examples include systems
of $N$ two-level systems ('atoms', either real or artificial), coupled
to a photon mode in a cavity~\cite{Kaluzny1983,Wallraff2004}. In this
case, time dependence might be introduced by changes of either the
photon frequency (by changing the cavity's size), or the energy
splitting of the two level systems (by applying a 'Zeeman' field.)
Similar physics also arises in the context of polaritons, excitons
coupled to a cavity electromagnetic mode~\cite{Weisbuch1992}. Another
phenomenon relevant to the present work is the observation of molecule
production in an atomic gas experiment, due to sweeping through a
Feshbach resonance \cite{Jin2004,Ketterle2004}.  While the fast sweep
regime was analyzed in Refs.~\cite{Altman2005,Barankov2005}, the
complementary case of slow sweeping rates, equivalent to a many-body
LZ problem~\cite{Pokrovsky2006,Pazy2006}, has not yet been understood.

Having the above setup of two level systems coupled to a
cavity  mode in mind, we consider the Hamiltonian
\begin{equation} \label{eq:ham}
H = -  \lambda t ~\hat b^\dagger \hat b + \frac{\lambda t} 2 \sum_{i=1}^N
\sigma^z_i + \frac{g}{\sqrt{N}} \sum_{i=1}^N \left( \hat b^\dagger \,
  \sigma^-_i + \hat b \, \sigma^+_i \right),
\end{equation}
where $\hat b^\dagger$ creates a photon mode, and $\sigma^\pm_i$ are
raising and lowering operators of the $i-$th two level
system. ($\sigma^\pm \equiv (\sigma^x \pm i \sigma^y)/2$, where
$\sigma^{x,y,z}$ are Pauli matrices.) The energy of the photon and the
two-level system vary in time as $\pm \lambda t$, respectively. The
Hamiltonian (\ref{eq:ham}) is equivalent, up to a gauge
transformation, to $H = - 2 \lambda t ~\hat b^\dagger \hat b + \omega_0 \sum_{i=1}^N \sigma^z_i + \frac{g}{\sqrt{N}} \sum_{i=1}^N \left( \hat b^\dagger \, \sigma^-_i + \hat b \, \sigma^+_i \right)$,
which represents a generalization of the James-Cumming
Hamiltonian~\cite{Jaynes1963} to $N$ two-level systems.  Equivalently,
we can think of (\ref{eq:ham}) as an effective Hamiltonian describing
a Feshbach resonance scenario: representing the
spin operators in (\ref{eq:ham}) in terms of Anderson pseudospin
operators,
\begin{equation} \sigma^z_i \rightarrow \hat a^\dagger_{i\uparrow}
  \hat a_{i \uparrow} - \hat a_{i \downarrow} \hat
  a^\dagger_{i\downarrow} , \, \sigma^+_i \rightarrow \hat
  a^\dagger_{i \uparrow} \hat a^\dagger_{i \downarrow}, \,
  \sigma^{-}_i \rightarrow \hat a_{i\downarrow} \hat
  a_{i\uparrow}
\end{equation}
where $\hat a^\dagger_{i\uparrow}$, $\hat a^\dagger_{i\downarrow}$,
$\hat a_{i\uparrow}$, $\hat a_{i\downarrow}$ are the creation and
annihilation operator for the spin-$1/2$ fermions labeled by $i$,
Eq.~(\ref{eq:ham}) assumes the form (up to an unimportant constant)
\begin{eqnarray} \label{eq:ham2}
H = - \lambda t ~\hat b^\dagger \hat
  b + \frac{\lambda t} 2 \sum_{i=1}^N \left( \hat
    a^\dagger_{i\uparrow} \hat a_{i\uparrow} + \hat
    a^\dagger_{i\downarrow} \hat a_{i\downarrow} \right)+ \cr +
  \frac{g}{\sqrt{N}} \sum_{i=1}^N \left( \hat b^\dagger \, \hat
    a_{i\downarrow} \hat a_{i\uparrow} + \hat b \, \hat
    a^\dagger_{i\uparrow} \hat a^\dagger_{i\downarrow}\right).
\end{eqnarray}
This is nothing but the Hamiltonian describing the creation of
molecules out of $N$ fermion pairs in a Feshbach resonance
experiment~\cite{Pokrovsky2006,Barankov2005} (although the single mode
approximation, i.e. the neglect of bosonic dispersion, may be
problematic in that case).

Assuming that the boson level is initially empty, and all fermions
resident in the upper state (on account of the energy balance at large
negative times),
 \begin{equation}\label{eq:ini}
\VEV{\hat b^\dagger \hat b}=0,\qquad \VEV{\sigma^z_i} =1,
\;\;i=1,\dots,N,
\end{equation}
our goal is to compute the asymptotic distribution
\begin{equation}
n_b(t)=\VEV{\hat b^\dagger(t) \hat b(t)}
\end{equation} at $t \rightarrow \infty$, i.e. the generalization of
the LZ transition probability $P$.
For $N=1$, this task is equivalent to the standard LZ problem, whose
answer is given by (cf. Eq.~\rf{eq:mLZ}) 
  $\lim_{t \rightarrow \infty} n_b(t) = 1-e^{-\frac{\pi
      g^2}{\lambda}}$.
However, for $N>1$, \rfs{eq:ham} defines a genuine
many-body problem and the solution of the corresponding
Schr\"odinger equation becomes progressively more difficult.

While we do not know how to handle the problem for arbitrary $N$, an
approximate solution valid in the limit of large particle numbers can
be found. At large $N$, the number of produced bosons turns
out to be reasonably well approximated by
\begin{eqnarray} \label{eq:mbLZ}
\lim_{t \rightarrow \infty}n_b(t) = & e^{\frac{\pi g^2}{\lambda}}-1,
&\qquad e^{\frac{\pi g^2}{\lambda}} \ll N, \\ \label{eq:mbLZ1}
\lim_{t \rightarrow \infty}n_b(t) \sim & \frac{ e^{\frac{\pi g^2}{\lambda}}}{\frac  2 N
e^{\frac{\pi g^2}{\lambda}} +1 }, &
\qquad e^{\frac{\pi g^2}{\lambda}} \sim N, \label{eq:mbLZ2} \\
\lim_{t \rightarrow \infty} n_b(t)  \rightarrow & N, & \qquad e^{\frac{\pi g^2}{\lambda}} \gg N.
 \label{eq:mbLZ3}
\end{eqnarray}

According to these equations, the adiabatic ground state
($n_b\stackrel{t\to\infty}{\longrightarrow} N$) is only realized if $
\lambda \ll \pi g^2/\log(N)$, a criterion which is progressively more
difficult to satisfy as $N$ becomes larger (see Ref.~\cite{Polkovnikov2007} for a general
discussion of the applicability of the adiabatic limit in large systems). This is in marked contrast
to the few body case, where adiabatic ground state occupancy is
granted for large values of the LZ parameter $e^{\frac{\pi
    g^2}{\lambda}}$. The observation of this
difference, obtained for a specific
model but likely indicative of a more general phenomenon, represents
the main result of the paper.

Technically, Eqs.~\rf{eq:mbLZ} and  Eq.~\rf{eq:mbLZ1}
obtain by integration of a rate equation for the boson occupation
number. Denoting the occupation of individual fermion states by $n_f$,
the latter reads
\begin{eqnarray} \label{eq:kinetic}
&&\d_t n_b =  2 \pi g^2 \delta(2 \lambda t) \left(
n_f^2 \left(1+n_b \right) - n_b \left(1-n_f \right)^2 \right),\nonumber\\
&&n_b + N n_f = N,
\end{eqnarray} 
where the factor $\delta(2\lambda t)$ accounts the energy balance in
particle conversion processes, the first (second) term on the right
hand side describes the creation (destruction) of bosonic particles by
destruction (creation) of two fermions, and the second line enforces
particle number conservation. Postponing the derivation of
Eq.~(\ref{eq:kinetic}), and the discussion of its range of validity to
below, we note that upon introduction of a variable $\theta$, such
that $\theta'(t)= \delta(t)$, Eq.~(\ref{eq:kinetic}) assumes the form
\begin{equation}
\d_\theta n_b = \frac{\pi g^2}{\lambda} \left(
n_f^2 \left(1+n_b \right) - n_b \left(1-n_f \right)^2 \right).
\end{equation}
At $\theta=1$ (which corresponds to $t \rightarrow \infty$) the
solution of this equation (with boundary condition $n_b=0$ at
$\theta=0$) reads
\begin{equation} \label{eq:sol}
n_b = \frac{ N \left(e^{ \frac{\pi g^2}{\lambda}}-1  
\right)}{   2 e^{ \frac{\pi g^2}{\lambda}}
+N },
\end{equation}
where terms of $\mathcal{O}(N^{-1})$ have been ignored.  Taking the
limit $N \rightarrow \infty$ at fixed $e^{\pi g^2/\lambda}$ gives
\rfs{eq:mbLZ}, while $e^{\pi
  g^2/\lambda} \sim N$ leads to \rfs{eq:mbLZ1}. Although that latter
limit is beyond the scope of
the large $N$ expansion, 
(\ref{eq:mbLZ1}) turns out to provide a reasonable (if uncontrolled)
approximation to $n_b$.

To actually derive \rfs{eq:kinetic}, we apply the Keldysh formalism.
Defining $t\equiv (t_1+t_2)/2$ and $\tau \equiv t_1-t_2$, we denote by
$G^{R,A,K}(t_1,t_2)\equiv G^{R,A,K}(t,\tau)$ the retarded, advanced,
and Keldysh fermionic propagators, respectively. (For the general
definition of Keldysh Green functions and notation conventions we
refer to the review~\cite{Kamenev2004}.) Initially all the $N$
fermionic levels are occupied; this corresponds to the bare
(noninteracting) propagators
\begin{equation} \label{eq:fermgreenfree}
G^{R,A}_0(t,\tau) = \mp i \theta(\pm \tau) e^{-i\frac{ \lambda }{2}  t \tau}, \ 
G^K_0(t,\tau)= i e^{-i \frac{\lambda}{2} t \tau},
\end{equation}
where the upper or lower sign in $\pm$ and $\mp$ are chosen for
retarded and advanced propagators respectively.  We aim to compute the
boson's Keldysh propagator $D^K(t,\tau)$ which, when evaluated at
$t=+\infty$, gives the number of produced bosons. Initially, however,
the boson level was unoccupied. Thus
\begin{equation}
D^{R,A}_0(t,\tau) = \mp \theta(\pm \tau) e^{i \lambda t \tau}, \  D^K_0(t,\tau) = -i e^{i \lambda t \tau}.
\end{equation} 
If the self energy $\Sigma(t,\tau)$ of the bosons is known, the
Keldysh bosonic propagator can be found by solving the 
Dyson equation, $(D_0^{-1} - \Sigma)\circ D=\openone$ where $D_0\,
(D)$ is the bare (dressed) bosonic propagator.  Introducing the
bosonic distribution matrix $F(t,t')$ through~\cite{Kamenev2004} $D^K
= D^R \circ F - F \circ D^A$, where $(A \circ B) (t_1,t_2)\equiv 
\int_{-\infty}^\infty dt_3\, A(t_1,t_3) B(t_3,t_2)$, and $D^{R,A,K}$
are the retarded, advanced and Keldysh components of $D$, the Dyson
equation for $D^K$ translates to a kinetic equation
\begin{equation} \label{kin}
\left[ F\stackrel{\circ}{,} i \partial_t + \lambda t \right] = \Sigma^K -
\left(\Sigma^R \circ F - F \circ \Sigma^A \right).
\end{equation}
\begin{figure}[hbt]
\includegraphics[height=.9 in]{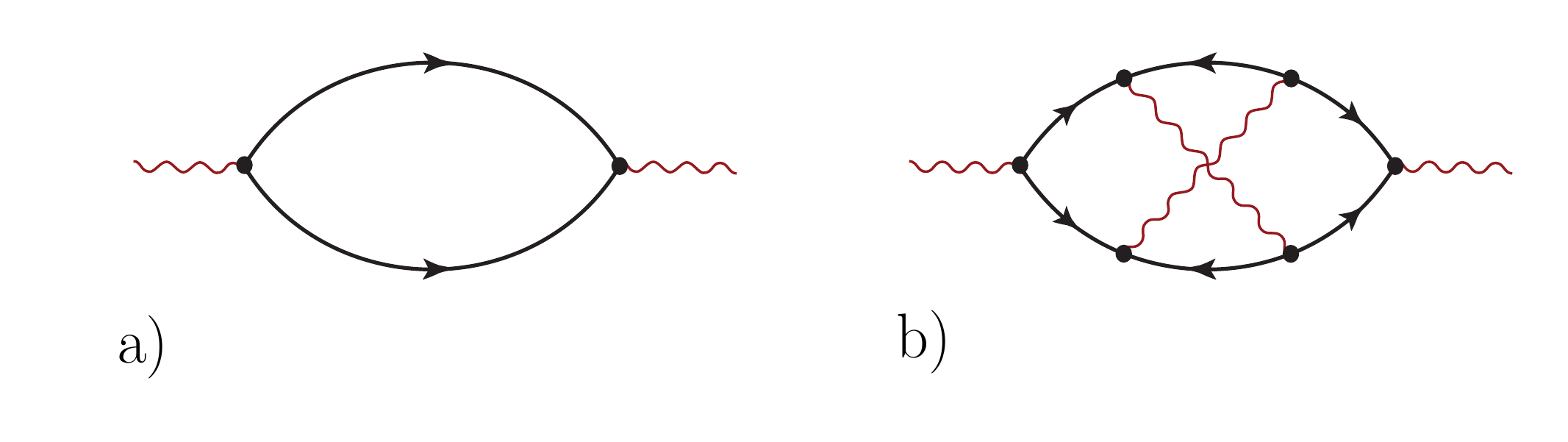}
 \caption{a) dominant self energy
  diagram for the bosonic propagator in the limit $N \rightarrow \infty$ with $g$ kept fixed. The straight lines are fermions,
  while the wavy lines are bosons. b) non-RPA diagram of lesser order
  in $N$.}
 \label{fig1}
\end{figure}
To approximately solve this equation, we note that only interaction
vertices $\sim g^2/N$ accompanied by one summation over $N$ fermion
states survive the limit $N\to \infty$. In practice, this means that
only the self energy diagram depicted in Fig.~\ref{fig1} a)
contributes to the boson self energy.  Processes such as the one shown
on Fig~\ref{fig1} b) are frustrated in that the number of fermion
summations does not compensate for the number of interaction
lines. One may verify that the same logics excludes any diagram other
than the one shown on Fig.~\ref{fig1} a). (For a caveat in the
argument, see below.)

The diagram shown in Fig.~\ref{fig1} a) translates to 
\begin{eqnarray}\label{se1} \Sigma^{R,A}(t,t') &=& i g^2 \int \frac{d\omega}{2\pi} \, G^{R,A}\left(t,\epsilon-\omega \right) G^K\left(t, \omega\right) \\ \nonumber
 \Sigma^K(t,t') &=& i \frac{g^2}{2} \sum_{k=R,A,K} \int \frac{d\omega}{2\pi} \, G^K\left(t,\epsilon-\omega \right) G^K\left(t, \omega\right).
\end{eqnarray}
on the right hand side we have switched to a Wigner representation,
\begin{equation}
G(t,\epsilon) \equiv  \int_{-\infty}^\infty d\tau~G(t,\tau) ~e^{i \tau \epsilon}.
\end{equation}
Introducing the spectral function
\begin{equation} \label{eq:deltaf}
\Delta_f(t,\epsilon)= 2 ~{\rm Im}~G^R(t,\epsilon)
\end{equation}
we obtain an equation for the Wigner transform of $F$, 
\begin{eqnarray} \label{eq:kintrans}
\left( \d_t-\lambda \d_\epsilon \right) F(\epsilon) = \frac{g^2}{2}
\int \frac{d\omega}{2\pi} \Delta_f \left(\epsilon-\omega \right) 
\Delta_f \left(\omega \right) \left[
1+ \right. \cr \left. f\left(\epsilon-\omega \right) f\left(\omega \right) - \left(f
\left(\omega \right)+f\left(
\epsilon-\omega \right) \right) F\left(\epsilon \right) \right].
\end{eqnarray}
Here $f(t,\epsilon)$ is the fermionic distribution function, and the
argument $t$, identically carried by all Wigner functions, is
suppressed for brevity.  In deriving \rfs{eq:kintrans} we assumed that
the Wigner transform of products of operators on the right hand side
(e.g. $(\Sigma^R\circ F)(\epsilon,T)$) can be replaced by the product
of the Wigner functions ($\Sigma(\epsilon,T) F(\epsilon,T)$). As
discussed a few paragraphs further down, this leading adiabatic
approximation~\cite{Kamenev2004} turns out to be exact in our
case. Also note that \rfs{eq:kintrans} was derived without specifying
whether the fermionic propagators in Fig.~\ref{fig1} a) are bare or
dressed.

Noting that in the distant future fermions and bosons become
effectively uncorrelated and the energy of the latter asymptotes to
$\epsilon=-\lambda t$, our aim is to calculate the bosonic
distribution function, $n_b(t) \equiv n_b(t,\epsilon=- \lambda
t)$. To transform Eq.~(\ref{eq:kintrans}) into an equation for $n_b$
we use the general relations $n_b = (F-1)/2$, $n_f=(1-f)/2$, and note
that $d_t n_b = (\partial_t - \lambda \partial_\epsilon)F/2$.
Approximating the fermion spectral functions by their bare value,
$A(\epsilon)=-2\pi \delta(\epsilon-\lambda t)$, we then readily arrive
at Eq.~(\ref{eq:kinetic}), where all fermionic distribution functions
$n_f(t) \equiv n_f(t,0)$ are evaluated at zero energy.   

Let us examine the status of the various approximations used in the
derivation of the rate equation: In the language of diagrammatic
perturbation theory, Eq.~(\ref{eq:kintrans}) treats the bosonic and
fermionic Keldysh components of the self energy operators in a self
consistent RPA approximation (i.e. a scheme wherein all 'crossing'
interaction lines in the bosonic and fermionic self energy are
neglected, on account of the limit $N\to \infty$.)  Notice that a {\it
  naive} interpretation of the large $N$ limit would suggest to
neglect the fermion self energy altogether: interaction corrections to
the fermion propagator do not come with a final state summation and
are, therefore, superficially of $\mathcal{O}(N^{-1})$. However, this
argument neglects that in regimes (\ref{eq:mbLZ1}) and
(\ref{eq:mbLZ3}) above, the bosonic distribution function $F\sim
n_b\sim N$ introduces additional $N$ dependence into the
theory. (Physically, the macroscopic occupation of the boson level
effectively enhances the fermion scattering rate.)  This mechanism
requires us to keep the RPA self energy of the fermionic {\it Keldysh}
Green function. However, the self energy corrections to the retarded
and advanced fermion propagators, which are independent of the bosonic
distribution function, are indeed large $N$ negligible. This latter
simplification justifies the above approximation of the fermion
spectral function by its bare value. (For the sake of completeness we
mention the existence of {\it non}-RPA diagrams in which a nominally
low power in $N^{-1}$ competes with factors $n_b$. [This happens,
e.g., in the Keldysh sector of the diagram shown in Fig.~\ref{fig1}
b).] These processes are not captured in our present analysis which
means that the theory becomes effectively uncontrolled once $n_b \sim
N$.)

We finally comment on the status of the leading order Moyal expansion
$(\Sigma \circ F)(\epsilon,T) \simeq \Sigma(\epsilon,T) F(\epsilon,T)$
used in the derivation. The temporal singularity $\sim \delta(t)$ of
the collision integral makes one worry that this replacement may,
indeed, not be entirely innocent. While we cannot really justify the
approximation in the resonant time window $t\sim 0$, we have checked
that it does yield the correct long time asymptotics (\ref{eq:mLZ})
when applied to the \textit{standard} LZ evolution equation.

It is instructive to reconsider the derivation of Eq.~(\ref{eq:mbLZ})
from a somewhat different perspective: the fact that the Hamiltonian
(\ref{eq:ham}) contains the Pauli matrices only in certain linear
combinations enables us to attack the problem by spin algebraic
methods. We define an ${\rm SU}(2)$ algebra of spin operators
$\{S^z,S^+,S^-\}$ acting in an $N/2$ dimensional Hilbert space as
$\hat S^z=\oh \sum_{i=1}^N \sigma^z_i$, $\hat S^\pm=\sum_{i=1}^N
\sigma^\pm$. Eq.~(\ref{eq:ini}) enforces full
initial polarization, $\VEV{\hat S^z} = N/2$.

Since the total number of bosons produced is much less than $N$ (the
defining criterion of the regime \rfs{eq:mbLZ}), the spin will not
deviate much from the vertical direction, and it is convenient to
employ a Holstein-Primakoff representation: replacing~\cite{Beige2005}
$\hat S^+ \rightarrow \sqrt{N}\, \hat b_{HP}$, $\hat S^- \rightarrow
\sqrt{N}\, \hat b^\dagger_{HP}$, $\hat S^z = N/2 -\hat b^\dagger_{HP}
\hat b_{HP}$, where $\hat b_{HP}^\dagger$ and $\hat b_{HP}$ are the
creation and annihilation operators of an auxiliary
Holstein-Primakoff boson, the large $N$ limit of the Hamiltonian 
\rfs{eq:ham} reduces to the quadratic form
\begin{equation}\label{eq:hamHP}
H = - \lambda t ~\hat b^\dagger \hat b - \lambda t ~\hat b_{HP}^\dagger \hat b_{HP}  + g  \left( \hat b^\dagger \, \hat b^\dagger_{HP} + \hat b \, \hat b_{HP} \right).
\end{equation}
The solution of the equations of motion of (\ref{eq:hamHP}) then leads
to Eq.~(\ref{eq:mbLZ}). (In a slightly different context, these
equations have been solved in \cite{Kayali2003}, where \rfs{eq:mbLZ} was
derived for the first time.) However, the above method does not appear
to be straightforwardly extensible to the regime of large transition
rates,  \rfs{eq:mbLZ1}.

\begin{figure}[h]
  \includegraphics[height=1.8 in]{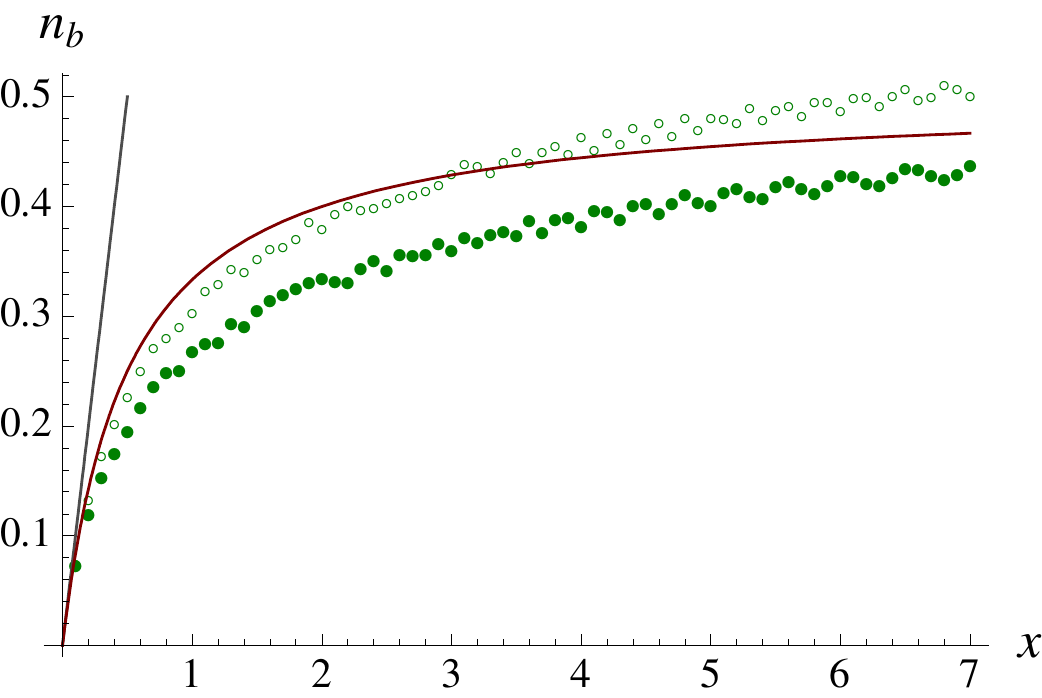} \caption{The boson
    production $n_b$ as a function of $x=\exp\left(\pi
        g^2/\lambda\right)/N$. Here $N=100$ (empty circles) and $N=500$ (filled circles). The straight line represents
    \rfs{eq:mbLZ} and the curve is
    \rfs{eq:mbLZ1}. The data is obtained by solving the Schr\"odinger equation for \rfs{eq:ham} at $g=1$, on the interval $-40\le t \le 40$, with the 
small oscillations in the data being the artifact of the finite time interval.}
 \label{fig3}
\end{figure}

To check the validity of our results we have run a numerical test. The
above spin representation shows that the Hilbert space of the problem
is of dimension $N+1$ (much lower than the $\mathcal{O}(2^N)$ naively
suggested by the representation (\ref{eq:ham})); this makes a
numerical solution of the Schr\"odinger equation
feasible. Fig.~\ref{fig3} shows $n_b$ as a function of
$x=\exp\left(\pi g^2/\lambda\right)/N$ for $N=100$ and $N=500$. At
$N=100$ the data is in general agreement with Eqs.~\rf{eq:mbLZ} and
\rf{eq:mbLZ1}, at larger $N$ we observe gradual downward deviations.
Preliminary results based on a combination of semiclassical ideas and
numerical integration indeed suggest the existence of corrections in
$\ln(N)$ (at fixed value $x >1$). However, in all our simulations, the
fraction $n_b/N$ converged to values {\it below} that predicted by
Eq.~\rf{eq:mbLZ1}, i.e. our principal observation of incomplete ground state
occupation remains valid.

To conclude, we have studied a  genuine many body generalization of the
Landau-Zener problem. Unlike with the single particle case, the system
does not settle in its many particle ground state and a finite fraction
of particles remains in energetically high-lying states. This
phenomenon makes many body Landau-Zener
physics profoundly different from the few body  case.

V.G. is grateful to the participants of the KITP, Santa Barbara,
program 'Strongly Correlated States in Condensed Matter and Atomic
Physics' and especially to R. Barankov, J. Keeling, A. Beige, 
V. Pokrovsky, and A. Polkovnikov for many comments and useful
discussions. A.A. acknowledges discussions with A.~Rosch and
B.~Simons. This work was supported by SFB/TR 12 of the Deutsche
Forschungsgemeinschaft and by NSF via grants DMR-0449521 and
PHY-0551164.


\end{document}